\documentclass[structabstract]{aa}  
\usepackage{graphicx}
\usepackage{times}
\usepackage{epsfig}
\usepackage{amsmath}
\usepackage{natbib}
\def\mpc{h^{-1}{\rm{Mpc}}}
\def\apj {ApJ}
\def\apjl {ApJL}
\def\apjs {ApJS}
\def\aj {AJ}

\def\aap {A\&A}
\def\mnras {MNRAS}

\def\etal{{\rm et al. }}
\def\mpc{{\  h^{-1} \rm Mpc}}
\def\kpc{{\ h^{-1} \ \rm kpc}}
\def\kms{{\ \rm km \ s^{-1}}}
\def\grad{^{\circ}}
\begin{document}

\title{  The orientation of galaxy pairs with filamentary structures: dependence on morphology }   

   \author{Valeria Mesa\inst{1}
   		  \and
          Fernanda Duplancic\inst{2}
          \and
          Sol Alonso\inst{2}
          \and
          M. Rosa Mu\~noz Jofr\'e\inst{3}
          \and
          Georgina Coldwell\inst{2}
          \and
          Diego G. Lambas\inst{3}
          }

   \institute{ Instituto Argentino de Nivolog\'{i}a, Glaciolog\'{i}a y Ciencias Ambientales (IANIGLA-CCT Mendoza, CONICET), Parque Gral San Mart\'{i}n, CC 330, CP 5500, Mendoza, Argentina\\
              \email{vmesa@conicet-mendoza.gob.ar}
         \and
             Departamento de Geof\'{\i}sica y Astronom\'{\i}a, Facultad de Ciencias Exactas, F\'{\i}sicas y Naturales, UNSJ-CONICET, San Juan, Argentina \\
            \and
            IATE, CONICET, OAC, Universidad Nacional de C\'ordoba, Laprida 854,
X5000BGR, C\'ordoba, Argentina\\
}
             
   \date{Received xxx; accepted xxx}

  \abstract
   {}
   {With the aim of performing an analysis of the orientations of galaxy pair systems with respect to the underlying large-scale structure, we study the alignment between the axis connecting the pair galaxies and the host cosmic filament where the pair resides. In addition, we analyze the dependence of the amplitude of the alignment on the morphology of  pair members as well as filament properties. }
 {We build a galaxy pair catalog requiring $r_p < 100\kpc$  and 
$\Delta V < 500 \kms $ within redshift $z<0.1$ from the Sloan Digital Sky Survey (SDSS). 
We divided  the  galaxy  pair  catalog  taking  into  account  the morphological classification by defining three pair categories composed by elliptical-elliptical (E-E), elliptical-spiral (E-S) and spiral-spiral (S-S) galaxies.
We use a previously defined catalog of filaments obtained from SDSS and we select pairs located closer than $1\mpc$ from the filament spine, which are considered as members of filaments. For these pairs, we calculate the relative angle between the axis connecting each galaxy, and the direction defined by the spine of the  parent filament.  
}
   {We find a statistically significant alignment signal between the pair axes and the spine of the host filaments consistent with a relative excess of $ \sim$  15\% aligned pairs. 
We obtain that pairs composed by elliptical galaxies exhibit a stronger alignment, showing a higher alignment signal for pairs closer than 200 $\kpc$  to the filament spine.
In addition, we find that the aligned pairs are associated with luminous host filaments populated with a high fraction of elliptical galaxies.

The findings of this work show that large scale structures play a fundamental role in driving galactic anisotropic accretion as induced by  galaxy pairs exhibiting a preferred alignment along the filament direction.}
   {}

 \authorrunning{Mesa \etal} 
 \titlerunning{The orientation of galaxy pairs with filamentary structures}             

   \maketitle
%

\section{Introduction}

Clusters, filaments, sheets and voids are the building blocks of the cosmic web.
It is believed that galaxies in filaments represent about the half of the baryon mass in the universe \citep[][and references therein]{fil3}.
In addition, \cite{Libeskind18} performed a comparative analysis of twelve different methods devised to classify the cosmic web finding a good agree between the most of the thechinque with a mass fraction in filaments between $\approx 10\%$ and  $\approx 60\%.$
According to the models of hierarchical structure formation, galaxy clusters grow through repeated mergers with other groups and clusters of galaxies \citep{zel, kat, bond, jen, col}. These processes occur anisotropically along preferential directions, indicating that galaxy clusters are fed through filaments containing individual galaxies and galaxy systems \citep{kod,ebe,pim}. The distribution and abundance of filaments may affect the properties of galaxies inhabiting these structures  \citep[e.g.][]{f2,f3}. Also, the large scale environments have influence on the formation and evolution of dark matter, it is therefore important to understand the correlations between the properties of halos and the topology of the cosmic web, because it may gives us valuable information about the physics of galaxy formation. 

On this topic, using simulations \citet{zha} conducted an investigation on the spins of the dark matter halos and the direction of the cosmic filaments. The authors found that both, the spins and the main axes of halos in filaments with masses $M\leq 10^{13} M_{\odot}$, are preferably aligned with the direction of the filaments, while spins and the main axes of halos in sheets tend to remain parallel. Also the authors found that with increasing halo mass the major axis tends to be more strongly aligned with the direction of the filament, but the alignment between the halo spin and filament becomes weaker for higher halo mass. 
In the similar direction, \cite{chen} from the study of the high-resolution hydrodynamical cosmological simulations, found that the galaxy alignment signal along filaments increases significantly with the subhalo mass.
Moreover, \cite{Libeskind12} using dark matter cosmological simulation examined the large-scale orientation of substructures and haloes with respect to the cosmic web, finding that the orbital angular momentum of subhaloes tends to align with the intermediate eigenvector of the velocity shear tensor for all haloes in knots, filaments and sheets. In addition, \cite{tyl13} show that the spin axis of spiral galaxies is found to align with the host filament, and also the minor axes of ellipticals are found to be preferentially perpendicular to hosting filaments.  

Furthermore, there exists a relation between satellite galaxies and their large scale environment \citep[][and references therein]{shao}. For example, based on numerical simulations \citet{bar15} predict a statistical excess of satellite galaxies with main axis aligned in the direction of the central galaxy. Evidence of this relation can be seen in the satellite population of M31, suggesting that tidal effects may have played an important role on its evolution. 
On this line, by using simulations and observational data \citet{fil1} study the dependence on the alignment amplitude of satellite galaxies with respect to filaments,  finding a statistically significant alignment signal between satellite position and filament axis. Also, they show that this alignment depends on the color/luminosity of the system, then it is stronger when the primary and satellite galaxies are brighter.
In addition, the authors suggest that the alignment signal may be a consequence of the satellite accretion via streams along the direction of the filaments. 
In addition, \citet{guo} showed that the satellite luminosity function of galaxies in filaments is significantly higher than those of galaxies not in filaments. The authors also found that the filamentary structures can increase the abundance of the brightest satellites by a factor of $\approx$2,  and this is independent of the primary galaxy magnitude. 

One the observational side, the pioneer study of \citet{lam88} accounts for a preferential distribution of bright galaxies according to their environment. The authors studied a sample of bright galaxies in rich clusters finding that, at scales up to 15 $\mpc$, galaxy counts are consistently higher in the direction of the major axes of bright clusters. 
In this line, \citet{bin} found that the orientation of the galaxy distribution in two neighboring clusters tend to be similar and, in addition, the brightest cluster galaxies has a tendency to be aligned with the distribution of galaxies in the system.

On the other hand, \citet{don06} developed a study of a sample of luminous red galaxies (LRGS) extracted from the fourth release of Sloan Digital Sky Survey (SDSS) within the redshift range 0.4 $<z <$0.5. They found a clear sign of alignment between the orientations of the LRGS and the distribution of galaxies within 1.5 $\mpc$. This alignment effect is present only for red tracers while the orientation of the LRGS is anti-correlated with the population of blue neighboring galaxies. These results could indicate the existence of a preferential direction of accretion in clusters, which also promotes the orientation of the brightest galaxies in the system. \citet{zha1} show that the major axes of galaxies in filaments are preferentially aligned with directions of the filaments, while galaxies in sheets have its major axes parallel aligned to the plane of the sheets. The strength of this alignment signal is stronger for red central galaxies, in agreement with results found for dark matter halos in N-body simulations \citep{lib,fil1}, suggesting that central red galaxies are well aligned with their host halos. These results are consistent with the works of  \cite{hirv} and  \cite{f3} who find a preferential alignment of red galaxies with the axis of SDSS filaments in the catalog of \cite{fil}. 

Galaxy interactions play an important role on the formation of the galaxies, since affect almost every aspect of the evolution of these objects \citep{alo06, woods07, elli10, lam03,lam12, mesa}. The presence of a close galaxy companion drives a clear enhancement in galaxy morphological asymmetries, and this effect is statistically significant up to projected separations of at least 50 $\kpc$ \citep{patton16}. Galaxy mergers have a relevant impact on the star formation activity since can trigger starbursts, and affect the galaxy stellar mass function \citep{gama}. The large scale environment can also affect the properties of interacting galaxies in pairs. On this line, \cite{fil2} calculated the angle between the line connecting galaxies of a sample of pairs with separations up to 1$\mpc$ and the direction of its host filament. The authors found that loose pairs, (i.e. pairs with projected separations greater than 300 $\kpc$) have a clear signal of alignment, with at least 25\% excess of aligned pairs, when compared with a random distribution.  

Motivated by these results, in this work we study the properties of close pair galaxies and the relation of a preferred orientation of pairs with respect to the underlying larger structures in which they are immersed. This paper is structured as follows: Section 2 describes the data used in this work, a detailed description of the catalog of filaments, and the procedure used to construct the pair catalog, explaining the classification process of the sample. In Section 3 we performed an study of the relative orientation of pairs of galaxies and filaments. 
Finally in Section 4, we summarize our main conclusions. 

Throughout this paper we adopt a cosmological model characterized by the parameters $\Omega_m=0.3$, $\Omega_{\Lambda}=0.7$ and $H_0=100 \kms \rm Mpc ^{-1}$.


\section{Data}

All data used in this work has been extracted from the Sloan Digital Sky Survey \citep[SDSS;][]{sdss}, one of the most successful surveys in the history of astronomy. Over years of operations (SDSS-I, 2000-2005; SDSS-II, 2005-2008; SDSS-III, 2008-2014) SDSS data have been annually released to the scientific community. The latest generation of the SDSS data \citep[SDSS-IV, 2014-2020;][]{sdssiv} is extending precision cosmological measurements to a critical early phase of cosmic history (eBOSS), expanding its infrared spectroscopic survey of the Galaxy in the northern and southern hemispheres (APOGEE-2), and for the first time using the Sloan spectrograph to make spatially resolved maps of individual galaxies (MaNGA).

In the present work we consider spectroscopic data from SDSS Data Release 8 \cite[DR8;][]{dr8}. This is the first release of SDSS-III survey and contains all of the imaging data taken by the SDSS imaging camera (over $14.000$ sq. deg. of sky), as well as new spectra taken by the SDSS spectrograph during its last year of operations for the SEGUE-2 project. DR8 is cumulative and includes essentially all data from the previous releases, the main improvement of this release has been a reprocessing of all the imaging data, and the stellar spectroscopy has been re-analyzed with a new stellar parameters pipeline.

Based on this data \cite{fil} built a filaments catalog using the sample of spectroscopic galaxies compiled in \cite{tem1}. Also using SDSS-DR8 we built a catalog of close galaxy pairs to perform the study proposed in this work. In this section we describe both the filaments and galaxy pair catalogs.

\subsection{Catalog of filaments}
\label{filaments}

The filament catalog derived by \cite{fil} was used for the purpose of this analysis. These authors consider galaxies within the adjacent main area of the SDSS Legacy Survey in the redshift range $0.009\leq z \leq 0.155$, the lower limit was set in order to exclude the local supercluster, while the upper limits represent a distance of 450 h$^{-1}$ Mpc. To define the filamentary structure, the authors  used the galaxy sample presented by \cite{tem1} and use Cartesian coordinates based on the angular coordinates of SDSS $\eta$ and $\lambda$. These coordinates are calculated according to:

\begin{align*}
\rm x &= \rm -d_{gal} \rm \sin \lambda \\
\rm y &=  \rm d_{gal} \rm cos \lambda \rm cos \eta \\
\rm z &= \rm d_{gal} \rm cos \lambda  \rm sin \eta
\end{align*}

where $d_{gal}$ is the comoving distance suppressed of Finger of God effect for galaxies, these coordinates are expressed in units of $\mpc$. For a detailed description of galaxy sample see \cite{tem1}.\\

They implement a statistical Bisous model algorithm based on the construction of cylinders to define a piece of filament. These cylinders are used to locate the filamentous structure using a marked point process with interactions, called Bisous model \citep{bisous}. To form a filament, at least two (preferentially three or more) cylinders must be aligned and connected.

The radius of a filament is set as r =0.5 $\mpc$, which is consistent with the average radius of groups/clusters of galaxies. Also it has been shown that filaments of this scale may influence the formation and evolution of galaxies \citep{Smith2012,tem13}.
The filament finder is probabilistic and gives the probability in density and orientation fields of the filament. Using these two fields, the filament spine is defined as the set of points with a separation of roughly $\sim0.5\mpc$ that determine the axis of the filament. Under this procedure single filaments can be extracted from data. 

The model described previously lead to a minimum number density inside a filament of 6 galaxies within 0.5 $\mpc$ radius, 6-10 $\mpc$ length cylindrical volume. Studying the distribution of filament length, these authors demonstrate that the longest filaments reach the length of 60$\mpc$. The filaments contain 35-40\% of the total galaxy luminosity and cover approximately 5-8\% of the total volume, in agreement with N-body simulations and previous observational results (e.g. Libeskind et al. (2018).

The final catalog comprises 15421 filaments, composed by nearly 400000 galaxies, and taking into account the limits imposed in the catalog of pairs with $z<0.1$ we found almost 300000 galaxies in the filament catalog that satisfy this condition. For these filaments the authors list different parameters such as the number of galaxies, total luminosity, filament length, number of points in the filament spine with a spacing $\sim 0.5\mpc$, filament minimum coordinate, and range in $x,y,z$ axis. And for the galaxies that compose the filaments we find useful information, for instance co-moving distance of the galaxy, distance from the nearest filament axis, id of the nearest filament, id of the nearest filament point. With these data we will build the catalog of pairs belonging to filaments. More information about the filament sample can be found in \cite{fil}.

\subsection{Galaxy Pair Samples}

The data used to construct the galaxy pair sample were derived from the Main Galaxy Sample \citep[MGS;][]{mgs} obtained from the \texttt{fits} files at the SDSS home page\footnote{http://www.sdss3.org/dr8/spectro/spectro$\_$access.php}. For this sample, k-corrections band-shifted to $z=0.1$, were calculated using the software \texttt{k-correct\_v4.2} of \cite{kcorrect}. For the data set, k-corrected absolute magnitudes were calculated from Petrosian apparent magnitudes converted to the AB system.

To identify pairs we select galaxies with projected separation  $r_{p}< 100 \kpc$ and relative radial velocities $\Delta V < 500 \kms$, within $z<0.1$. With these restrictions we obtained an initial sample of 25965 galaxy pairs in the central area of the Legacy Survey of SDSS. To achieve the morphological classification we cross-correlated our sample of galaxy pairs with the Galaxy Zoo catalog\footnote{http://www.galaxyzoo.org/} \citep{zoo, zoo2} that comprises a categorization of nearly 900,000 galaxies drawn from the the spectroscopic data of the Sloan Digital Sky Survey.

\begin{figure*}
   \centering
 \centerline{\psfig{file=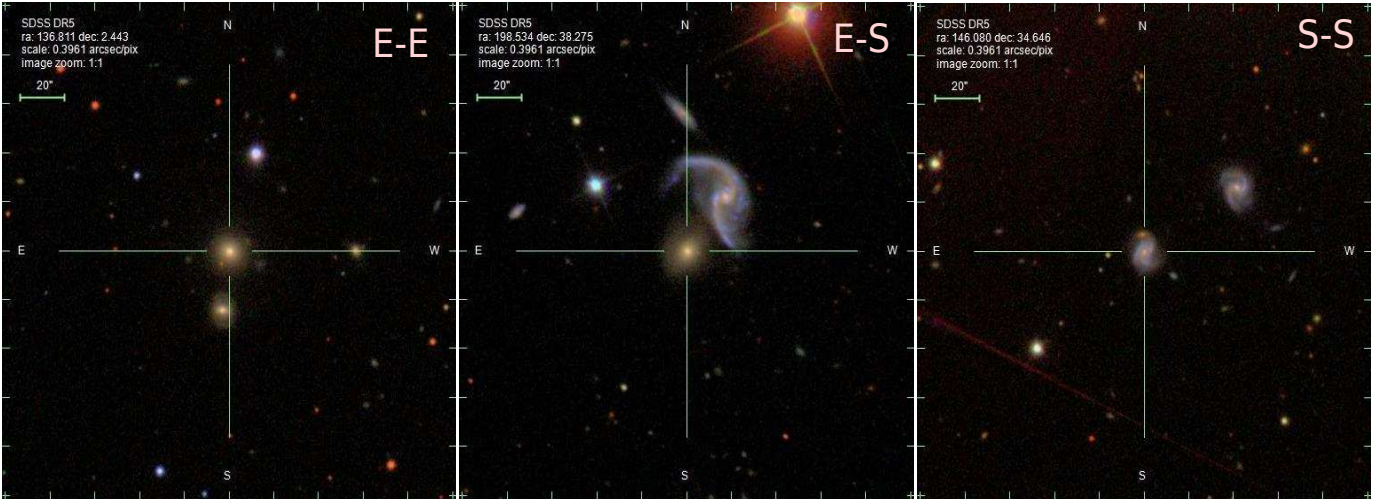,width=18cm,height=6.6cm}}
\caption{ Examples of galaxy pairs in E-E, E-S and S-S samples.  }
\label{ej1}
\end{figure*}

Regarding the morphological classification there are six categories, elliptical, spiral, spiral clockwise, spiral anticlockwise, merger or uncertain. For a given galaxy the catalog provides the fraction of votes. For our data we selected galaxies classified as spiral (S) or elliptical (E) considering a debiased vote fraction $>60\%$. 
We define three categories of the galaxy pairs; 1) pairs formed by two elliptical galaxies (E-E), 2) elliptical-spiral pairs (E-S) and 3) pairs formed by two galaxies with spiral morphology (S-S). Pairs not fulfilling these restrictions were excluded from the present study. The same classification scheme was adopted by \cite{iba} for a sample of isolated galaxy pairs to study nuclear activity as a function of morphology. Fig. \ref{ej1} shows typical examples of the three galaxy pair categories and Table\ref{tab2} provides the numbers and percentages of the pair samples defined previously in the central area of SDSS.

\section{Galaxy pairs in filaments}

\subsection{Correlation of galaxy pairs with the filament catalog}

In order to find galaxy pairs within filaments, we cross-correlated our sample of galaxies with the table of galaxies in filaments given by \cite{fil}, and we obtained for them the properties previously described in the section \ref{filaments}. Since pair members in our sample are close, we consider the position of the brightest member as the position of the pair. We take into account the filament thickness and we choose those pairs that are located at a smaller distance than $1\mpc$  from the nearest filament axis, for this purpose we use the distance provided in the filament catalog. 
With these restrictions, the pair sample associated to filaments is considerably reduced accounting for  $\sim35\%$ of the total sample. 

For these galaxy pairs within filaments we study the normalized distributions ($\rm N_{\rm bin}/\rm N_{\rm tot}$) of redshift ($z$) and r-band absolute magnitude ($M_r$), of every galaxy in the pairs, finding slight differences between E-E, E-S and S-S pairs. In addition, to avoid biases in the results we request all systems  to have comparable distances and luminosities, therefore we randomly select pairs to match redshift and magnitude distributions by using a Monte Carlo algorithm. This is accomplished by constraining simultaneously the two distributions so that each galaxy pair in the list is requested to fit the observed E-S distributions, since this sample has an intermediate behavior. Thus, E-E and S-S distributions, which have departures from the E-S sample, are pruned to behave in a similar fashion. 

\begin{figure} 
\centering
  \includegraphics[width=.50\textwidth]{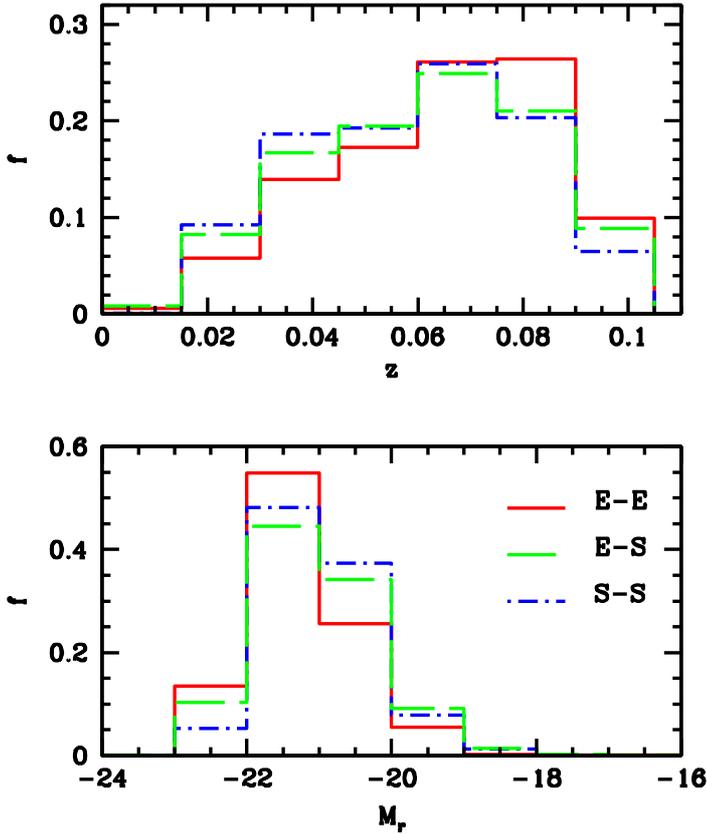}
\caption{Normalized distributions of $z$ and $M_r$ of E-E, E-S and S-S (solid, dashed and dot-dashed lines) pair galaxies associated to filaments.}
\label{z_r}
\end{figure}

Fig. \ref{z_r} shows the distributions of redshift, $z$, and absolute magnitude, $M_r$, for  the three samples of pairs in filaments studied in this work. In order to test the similarity between these distributions we performed Kolmogorov-Smirnov tests finding in all cases $p>0.5$. In this way, we have confidence that the three samples have similar distributions for these parameters.

It is worth to mention that the selected systems are embedded in filaments and their properties are likely to be affected by the large scale environment. Table \ref{tab2} provides the classification, number of pairs and percentages in these E-E, E-S and S-S pair samples associated to filaments. As it can be observed, the number of E-E its slightly lower than S-S pairs, however E-S pairs represent approximately the half of the total sample,  indicating that they are the most frequent combination of galaxies in pair systems inhabiting filaments. Noticeably, the environment plays a fundamental role since it favors the type of mixed interactions (E-S) in our sample of pairs immersed in dense structures. 

\begin{figure} 
\centering
 \includegraphics[width=.46\textwidth]{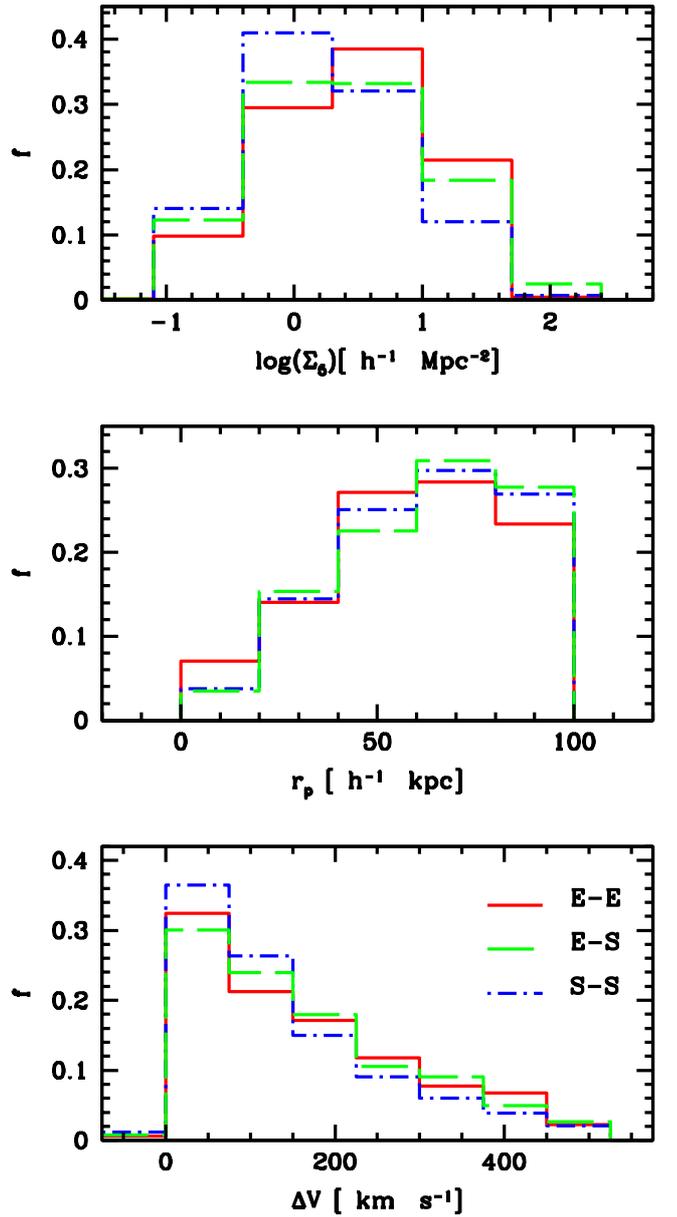}
\caption{Normalized distribution of $log(\Sigma_5)$, $r_p$ and $\Delta V$ of pair galaxies in E-E (solid), E-S (dashed) and S-S (dot-dashed) samples. }
\label{prop}
\end{figure}

In order to analyze the characteristics of the three galaxy pair samples we studied the distribution of the projected distance ($r_p$) and radial velocity difference ($\Delta V$) between galaxies in pairs, distinguishing between E-E, E-S and S-S samples. Furthermore we estimated the local density of every sample by using the local density estimator, $\Sigma_5$, in logarithmic scale.
The $\Sigma_5$ parameter\footnote{$\Sigma_5 = 5/(\pi d^2)$} is defined through the projected distance $d$ to the $5^{th}$ nearest neighbor brighter than $M_r < -20.5$ \citep{balo04}, (notice that this absolute magnitude threshold is consistent with our flux limited catalog  at the maximum distance of the pairs, $z< 0.1$), with a radial velocity difference less than 1000$\kms$, and provides a suitable measurement of the local density of the systems. Initially the samples showed slight differences in the environment they inhabited, for that reason we repeat the procedure explained above, adding this parameter to the fit, in order to obtain a sample independent of its environment. 
The results are shown in Fig \ref{prop} where, also we notice that the  mean separation between pair members $\sim 60 \kpc$ corresponds to an angular separation of $\sim 5$ arcmin at the mean redshift of the samples ($z\sim 0.06$) which allows to suitably determine the vector connecting pair members with SDSS astrometric precision. Velocity differences of the samples are more similar.

\begin{table}
\center
\caption{Classification, number of pairs and percentages in the E-E, E-S and S-S initial samples, and samples associated to filaments.}
\label{tab2} 
\begin{tabular}{| c c c | }
\hline\hline
 Classification & Number of pairs & Percentages\\
\hline
 E-E      &   2674   &    21.74\%  \\
 E-S      &   5162    &   41.97\% \\
 S-S      &   4462     &  36.29\% \\
 Total   &   12298     &   100\% \\
\hline
& \multicolumn{1}{c}{Pairs associated to filaments}\\
\hline
 E-E      &  995    & 21.56\%  \\
 E-S      &  2305      &   49.96\% \\
 S-S      &   1314        &       28.48\% \\
Total   &    4614     & 100\% \\           
\hline
\end{tabular}
\end{table}

\subsection{Relative orientation between galaxy pairs and the host filament}
\label{alpha}

This section provides a detailed analysis of the relative orientation of pairs formed by galaxies with different morphologies (E-E, E-S and S-S) with respect to their host filament. It should be taken into account that peculiar velocities produce redshift-space distortions on the radial component of galaxy positions, affecting distance estimators based on redshift. Therefore, in the forthcoming analysis we will consider distances projected in the plane of the sky. To define pair orientation we use the vector connecting galaxies of each pair and as tracer of the filament orientation we consider the id of the nearest filament point and we count five points towards each side of it, in such a way to consider an approximate radius of $2.5\mpc$ to trace a line connecting them. Then we measure the projected angle $\alpha$ between the filament orientation and pair orientation. In this way, we are considering filaments with a minimum length of $5\mpc$. A similar approach is used in \citet{fil2}.

We compute the distribution of relative angles F($\alpha$), defined as $F(\alpha)= (N(\alpha)-<N(\alpha)>)/<N(\alpha)>$, , 
where $N(\alpha)$ is the number of galaxy pairs with an angle $\alpha$ within each angular bin and $<N(\alpha)>$ is its mean value, that is, the expected value of pairs in that range, if the sample were uniform. This function is evaluated in a range $0\grad<\alpha<90\grad$. Then, if the pairs are aligned with the filaments, the distribution of $F(\alpha)$ will present an excess at low values of $\alpha$, a flat distribution of $F(\alpha)$ is consistent with a random distribution of relative angles. 

\begin{figure} 
\centerline{\psfig{file=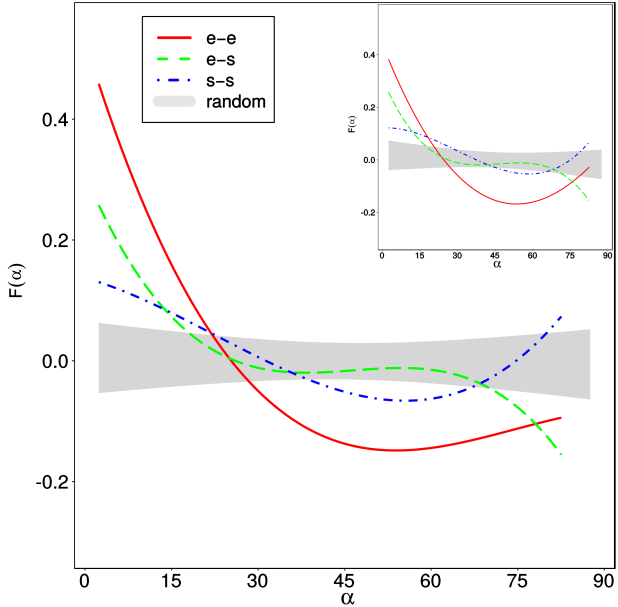,width=8.cm,height=8.cm}}
\caption{Relative fraction of galaxies $F(\alpha)$ for E-E pairs (solid line), E-S pairs (dashed line) and S-S pairs (dot-dashed line). The shadow area shows the 95\% confidence interval for a randomized distribution
with equal number of objects than the pair sample. Inbox: analogous analysis for a sample restricted to a similar environment.}
\label{fil2}
\end{figure}

In Fig \ref{fil2} we show the cubic smoothing spline distribution of F($\alpha$) for the angle between the orientation of E-E, E-S and S-S galaxy pairs and their host filament. We also plot $F(\alpha)$ for a sample of randomized angles with a number of points equal to the galaxy pair sample. For the aim of analysing if this effect is  independent of environment, the same analysis was also performed for a sample with a similar distribution in the $\Sigma_5$ parameter (see inbox Figure). As shown here, the alignment effect does not depend crucially on the local environment.  A similar kernel approach was used by \citet{fil2} to estimate alignment probability between pair orientation with filaments.
From this figure it can be seen that galaxy pairs tend to be aligned with their host filaments. There is an excess of pairs within 30 degrees along with a clear decrease of the profile at pair relative orientations perpendicular to the filament spine. In the case of E-E pairs the detection significance of the alignment signal for $\alpha < 20\grad$ is at $5 \sigma$ level when compared to a random distribution.

In order to  quantify the amplitude of the alignment signals of our samples, we compute the ratio between the number of aligned  to antialigned pairs  $\beta$ = $ N_{(<45)}/N_{(>45)}$. 
$\beta$ is linearly related to the $b$ parameter of the cosine model  \footnote{$\beta =$ $1 + b*0.52$} adopted in many works eg. \citep{lam88,don06} and references therein.

Table \ref{t3} lists the obtained $\beta$ values for each sample. By inspection to this table it can be clearly seen the tendency of pairs composed by elliptical galaxies (i.e. E-E and E-S) to be aligned with the filaments. This systematic alignment is also observed in the S-S sample, but with a lower amplitude.

\begin{table}
\center
\caption{$\beta$ parameter for the E-E, E-S, S-S and random samples.}
\begin{tabular}{| c c c c| }
\hline\hline
 Sample  & $\beta$  &  $\beta_{close}$  & $\beta_{far}$\\
\hline
 E-E         & 1.17 & 1.29 & 1.05 \\
 E-S         & 1.14 & 1.26 & 1.01 \\
 S-S         & 1.09 & 1.03 & 1.18 \\
\hline
\end{tabular}
\label{t3}
\end{table}

We also argue that there could be a dependence of the alignment signal on the proximity to filament  spine. To explore this possibility we calculated the median value of the pair-host filament relative distance finding a similar value for the three samples E-E, E-S and S-S $d_m\sim 200 \kpc$. Then we computed the $\beta$ alignment parameter for pairs with relative distances to their host filament lesser (larger) than $d_m$ ($\beta_{close}$) and ($\beta_{far}$) respectively.

In Table \ref{t3} we also show these values. It can observed that the  alignment signal of the pairs show significant variations with the distance to the host filament. With a high alignment signal for E-E and E-S  pairs closer to the filament spine. However, for the S-S pairs the trend is opposite.

We argue that the strong alignment signal derived for close pairs and filaments highlights a fundamental role of large scale structure in driving accretion onto along the preferred directions traced by filaments. \\

\subsubsection{ Relation between filament properties and alignment signals}
\label{propfil}

The goal of this section is to analyze the relation between the  alignment signals of galaxy pairs and filaments and the overall properties of host filaments. To this end, we use filament properties extracted from the catalog of \cite{fil}. 
In particular we consider the number of galaxies in filaments and the total filament luminosity (in units of $10^{10}\rm h^{-2} \rm L_\odot$) calculated by considering galaxies located closer than $0.5\mpc$ from the filament axis. We have also taken into account the fraction of spiral and elliptical galaxies per filament. 
We cross-correlated the sample of galaxies of the filament catalog of \cite{fil} (descripted in section \ref{filaments}) with the Galaxy Zoo catalog. For each galaxy in this sample with an associated Id of a filament, it was assigned a morphological type given by Galaxy Zoo,  selecting galaxies classified as spiral (S) or elliptical (E) considering a debiased vote fraction $>60\%$. Then and we estimated the number of elliptical and spiral galaxies with respect to the total number of galaxies forming each filament.

\begin{figure} 
\centerline{\psfig{file=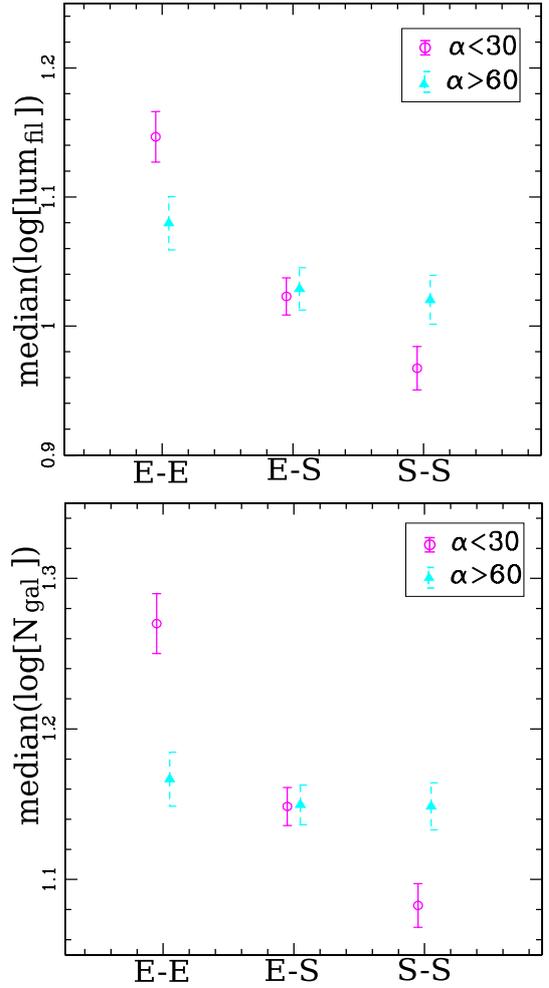,width=7cm,height=13cm}}
\caption{Top panel: Median luminosity (in units of $10^{10}\rm h^{-2} \rm L_\odot$), in logarithmic scale, of the host filaments of E-E, E-S and S-S pairs. Aligned pairs: empty circles, and antialigned pairs filled triangles. Bottom panel: Same as Top panel for the median of the number of galaxies. All the uncertainties  were derived through a bootstrap resampling technique \citep{boos}.}
\label{fil4}
\end{figure}

Figure \ref{fil4} shows the  median of the number of galaxies ($N_{gal}$) and total luminosities ($\rm lum_{\rm fil}$) distinguishing between filaments hosting E-E, E-S and S-S aligned pairs ($\alpha<30\grad$) and host filaments of antialigned systems ($\alpha > 60\grad$). We see that the aligned pairs show variations according to their morphology and to the properties of the host filament  with a decreasing trend of the median luminosity and number of galaxies of filaments hosting E-E to S-S pairs. On the other hand no significant dependence is observed for antialigned pairs which show a fairly uniform distribution within the estimated errors. 

E-E pairs, which reside in more luminous filaments and with a larger number of galaxies, show $\sim 10\%$ larger values of the median of $N_{gal}$ for filaments hosting aligned pairs with respect to antialigned systems. Otherwise, we find a marginal tendency for aligned S-S pairs to reside in less luminous filaments and with a lower number of galaxies than non-aligned systems.  We have also tested these trends in  samples restricted to a narrow redshif range finding a similar behavior. \\

\begin{figure} 
\centering
\centerline{\psfig{file=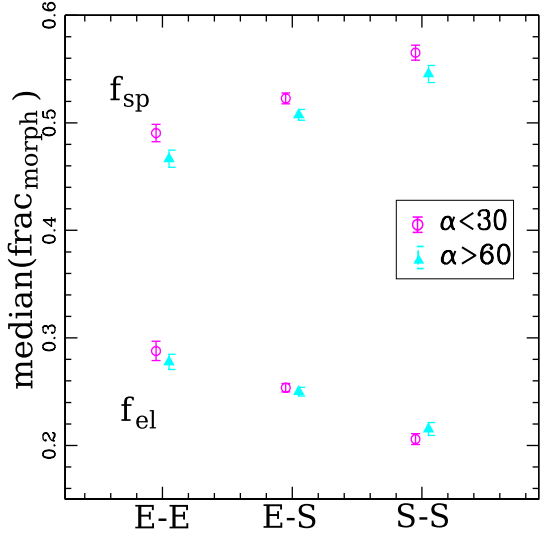,width=7.0cm,height=7.0cm}}
\caption{Median values of fractions of elliptical and spiral galaxies per filament, in E-E, E-S and S-S pair samples. Aligned pairs: empty circles, antialigned pairs: filled triangles. All the uncertainties  were derived through a bootstrap resampling technique.}
\label{fil5}
\end{figure}

Regarding the fraction of elliptical and spiral galaxies in each filament, we find that in all the samples the filaments are dominated by spiral galaxies, an expected result, given that spiral galaxies are the most abundant in the universe \citep{wil,tem0}.  Fig \ref{fil5} shows the median values of fractions of elliptical and spiral galaxies in filaments hosting E-E, E-S and S-S pairs. From this figure it can be seen that, as expected, there is a preference in the E-E pairs to reside in filaments with a larger fraction of elliptical galaxies, conversely, S-S pairs tend to reside in filaments populated mainly by spirals. Nevertheless, no significant differences are observed regarding the relative alignment of pairs and the host filament.

Summing up the results of this subsection we can conclude that S-S pairs reside preferentially in low luminosity filaments whereas E-E pairs tend to be more strongly aligned in filaments with larger number of members.
The fraction of spirals (ellipticals) in host filaments increase (decrease) from E-E to S-S pairs, as expected.

\section{Summary and Discussion}

We use a spectroscopic sample derived from SDSS and select galaxy pairs considering projected distances $r_p < 100 \kpc$ and radial velocity differences $\Delta V < 500 \kms$, within $z<0.1$. With the aim of understanding the impact of morphology in our studies, we have used the Galaxy Zoo catalog to divide the samples into pairs composed of two elliptical galaxies (E-E), one elliptical and one spiral (E-S) and two spiral galaxies (S-S). 

In order to study the presence of alignment effects with larger structures we use the filament catalog of \citet{fil} and select galaxy pairs located closer than 1$\mpc$ from the filament spine. To avoid the effect of the peculiar velocities on the results, we work in projection on the plane of the sky. We compute the angle $\alpha$ between the axis connecting pair members and the direction axis of filaments and we consider the distribution function  $F(\alpha)$ to measure the excess of pairs.
We measure the ratio $\beta$ of aligned to antialigned pairs as a suitable measure of the preferred orientation effect with typical values $\beta \sim 1.15$.  Finally we study the dependence of the alignment on  filament and galaxy properties

Our main results can be summarized as follow: 

\begin{itemize}

\item We determine that pairs composed by two elliptical galaxies tend to be strongly aligned with the parent filament spine. This tendency increases for pairs closer to the axis of the filament. 

\item Pairs composed by two spiral galaxies show a weaker alignment signal that increases for pairs at larger separations from the filament.

\item The global properties of the filaments affect significantly the pair alignment signal:

\item The number of galaxies in host filaments is higher for associated aligned E-E pairs compared to host filaments in antialigned E-E systems. 

\item On the other hand, aligned S-S pairs reside in less luminous filaments than non-aligned S-S systems.
\\ 

\end{itemize}

We obtain a significant dependence of the relative alignment of the pair orientations and nearby filaments on both galaxy morphology and distance to the filament. Close galaxy pairs show a preference to be aligned with the filamentary structures, particularly those formed by elliptical galaxies.  
 
In general, reported alignment signal in previous works have concerned  samples of ellipticals \citep{lam88,don06}. We stress the fact that here, samples of close pairs composed by spirals also exhibit significant alignment signals. 

The alignment of galaxy pairs and filaments was studied in \cite{fil2}, where the effects are detected for pairs of galaxies with relative  separations of the order of 1$\mpc$. 
In the present paper, this reported alignment signal extends to close galaxy pairs at significantly smaller relative separations $\sim 30 - 60  kpc$ which are near the process of undergoing a merger event. 

In this context, the findings of this study show evidence that the global environment has a key role in driving accretion of galaxy pairs with a preferred orientation along the filament direction.
Moreover, we find that the morphology of pair member galaxies  and global features of the cosmic filaments are important issues to take into account in this large-scale -- galaxy interplay.

\begin{acknowledgements}
We would like to thank to the referee Dr. Elmo Tempel for a detailed revision of the manuscript and for the suggestions that helped to improve this paper.    
This work was partially supported by the Consejo Nacional de Investigaciones
Cient\'{\i}ficas y T\'ecnicas and the Secretar\'{\i}a de Ciencia y T\'ecnica 
de la Universidad Nacional de San Juan.

Funding for SDSS-III has been provided by the Alfred P. Sloan Foundation, the Participating Institutions, the National Science Foundation, and the U.S. Department of Energy Office of Science. The SDSS-III web site is http://www.sdss3.org/.

SDSS-III is managed by the Astrophysical Research Consortium for the Participating Institutions of the SDSS-III Collaboration including the University of Arizona, the Brazilian Participation Group, Brookhaven National Laboratory, Carnegie Mellon University, University of Florida, the French Participation Group, the German Participation Group, Harvard University, the Instituto de Astrofisica de Canarias, the Michigan State/Notre Dame/JINA Participation Group, Johns Hopkins University, Lawrence Berkeley National Laboratory, Max Planck Institute for Astrophysics, Max Planck Institute for Extraterrestrial Physics, New Mexico State University, New York University, Ohio State University, Pennsylvania State University, University of Portsmouth, Princeton University, the Spanish Participation Group, University of Tokyo, University of Utah, Vanderbilt University, University of Virginia, University of Washington, and Yale University. 

\end{acknowledgements}
\bibliographystyle{aa}

\end{document}